\documentclass[twoside]{IEEEtran}
\usepackage{acronym}
\usepackage{psfig}
\usepackage{alg}
\usepackage{side}
\usepackage{times}
\usepackage{html}
\usepackage{fancyhead}
\usepackage{lgrind}

\ifx \htmlstyloaded\relax  \else\let\htmlstyloaded\relax\fi

\newcommand{\htmladdnormallink}[2]{#1}

\newcommand{\htmladdimg}[1]{}

\newcommand{\externallabels}[2]{}

\newcommand{\externalref}[1]{}

\makeatletter
\def\makeinnocent#1{\catcode`#1=12 }
\def\csarg#1#2{\expandafter#1\csname#2\endcsname}

\def\ThrowAwayComment#1{\begingroup
    \def\CurrentComment{#1}%
    \let\do\makeinnocent \dospecials
    \makeinnocent\^^L
    \endlinechar`\^^M \catcode`\^^M=12 \xComment}
{\catcode`\^^M=12 \endlinechar=-1 %
 \gdef\xComment#1^^M{\def\test{#1}
      \csarg\ifx{PlainEnd\CurrentComment Test}\test
          \let\html@next\endgroup
      \else \csarg\ifx{LaLaEnd\CurrentComment Test}\test
            \edef\html@next{\endgroup\noexpand\end{\CurrentComment}}
      \else \let\html@next\xComment
      \fi \fi \html@next}
}
\makeatother

\def\includecomment
 #1{\expandafter\def\csname#1\endcsname{}%
    \expandafter\def\csname end#1\endcsname{}}
\def\excludecomment
 #1{\expandafter\def\csname#1\endcsname{\ThrowAwayComment{#1}}%
    {\escapechar=-1\relax
     \csarg\xdef{PlainEnd#1Test}{\string\\end#1}%
     \csarg\xdef{LaLaEnd#1Test}{\string\\end\string\{#1\string\}}%
    }}

\excludecomment{comment}

\excludecomment{rawhtml}

\excludecomment{htmlonly}
\newcommand{\html}[1]{}

\newcommand{\hyperref}[4]{#2\ref{#4}#3}

\newcommand{\htmlimage}[1]{}

\newcommand{\htmladdtonavigation}[1]{}

\begin{document}
%
%
\title{Brittle System Analysis}
\author{\htmladdnormallink{Stephen F. Bush}
{http://www.crd.ge.com/people/bush}, John Hershey and Kirby Vosburgh\thanks{Stephen F. Bush, John Hershey and Kirby Vosburgh 
General Electric Corporate Research and Development, KWC-512, One Research
Circle, Niskayuna, NY 12309 e-mail: bushsf@crd.ge.com}}
\maketitle
\markboth{Bush, Hershey, and Vosburgh: Brittle Systems}
{Bush, Hershey, and Vosburgh: Brittle Systems}
\begin{abstract}

The goal of this paper is to define and analyze systems which
exhibit brittle behavior. This behavior is characterized by
a sudden and steep decline in performance as the system
state changes. This can be due to input parameters which
exceed a specified input, or environmental conditions
which exceed specified operating boundaries. 

\end{abstract}

\begin{keywords}
System Design, Catastrophe Theory, Quality
\end{keywords}

%
%
\section{Introduction}
\label{introduction}

\begin{figure*}[htbp]
        \centerline{\psfig{file=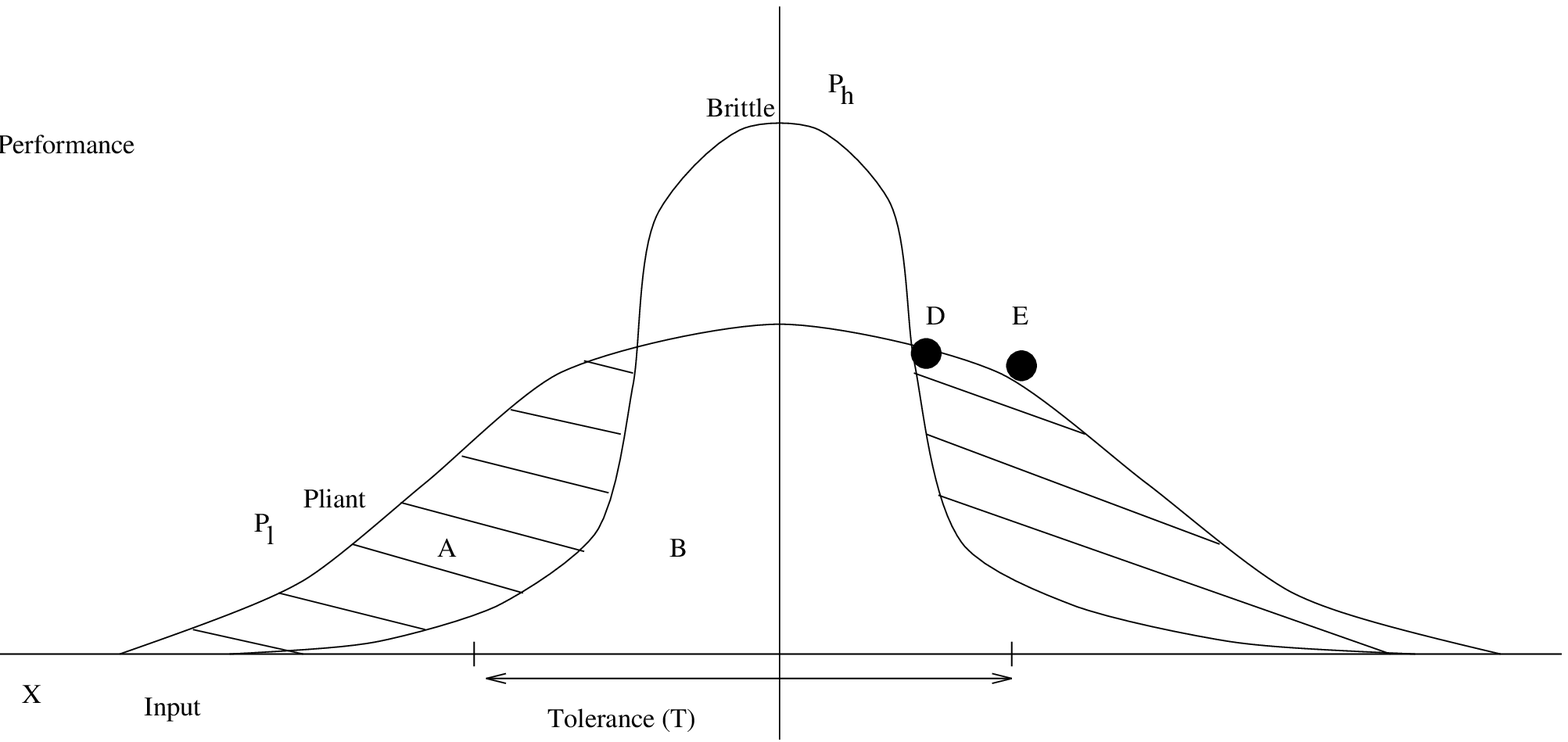,width=6.0in}}
        \caption{A Brittle versus Ductile System.}
        \label{bintro}
\end{figure*}

\PARstart{B}{ased} on vast experience watching the fruit of my hard
work fall apart time and again, 
I feel highly qualified to discuss the manner in which
systems break. In particular, the goal of this paper is to define and 
analyze systems which exhibit brittle behavior. This behavior is 
characterized by a sudden and steep decline in performance as system
state changes as shown by point D along curve $P_h$ in Figure \ref{bintro}.
$P_h$ is the performance curve for a high performance system with
brittle characteristics, $P_l$ is a lower performance system with less
brittle characteristics. Clearly the slope from point $D$ along curve
$P_h$ is much steeper than that of point $E$ along curve $P_l$. 
The steep decline of performance along $P_h$ can be due to input 
parameters which exceed a specified tolerance, 
or environmental conditions which exceed specified operating boundaries. 
This is equivalent to material fracture. Materials science 
provides a terminology which is apropos and flexible enough to describe
the characteristics of this work. A table of materials science terms and
their corresponding brittle system definitions is shown in Table
\ref{terms}. Toughness \cite{Vlack} is the amount of energy absorbed 
by a material prior to failure. A brittle fracture occurs with very little 
energy absorption while a ductile fracture is accompanied by much energy
absorption. Clearly toughness is the analog of the robustness of a system.
To carry the analogy further, ductility is quantified as the 
amount of permanent strain prior to fracture. A system which does not 
exhibit brittle behavior will be called ductile \cite{Vlack}. 
Strain is unit-less and refers to the amount of deformation per unit length 
of a material and is caused by stress which is the force per unit area. 
Material deformation is analogous to degradation in a brittle system.
In our work, stress is the distance by which a parameter exceeds its 
specified operating tolerance. 
There are two forms of strain, reversible and permanent. Reversible 
strain is called elastic strain and is characterized by Young's modulus: 
the ratio of the stress over the strain. Permanent strain leaves the 
shape of a material permanently changed and is known as plastic strain. 
In a brittle system, plastic strain will be degradation from which the
system cannot recover, while a brittle system can recover from reversible 
strain.

\begin{figure*}[htbp]
        \centerline{\psfig{file=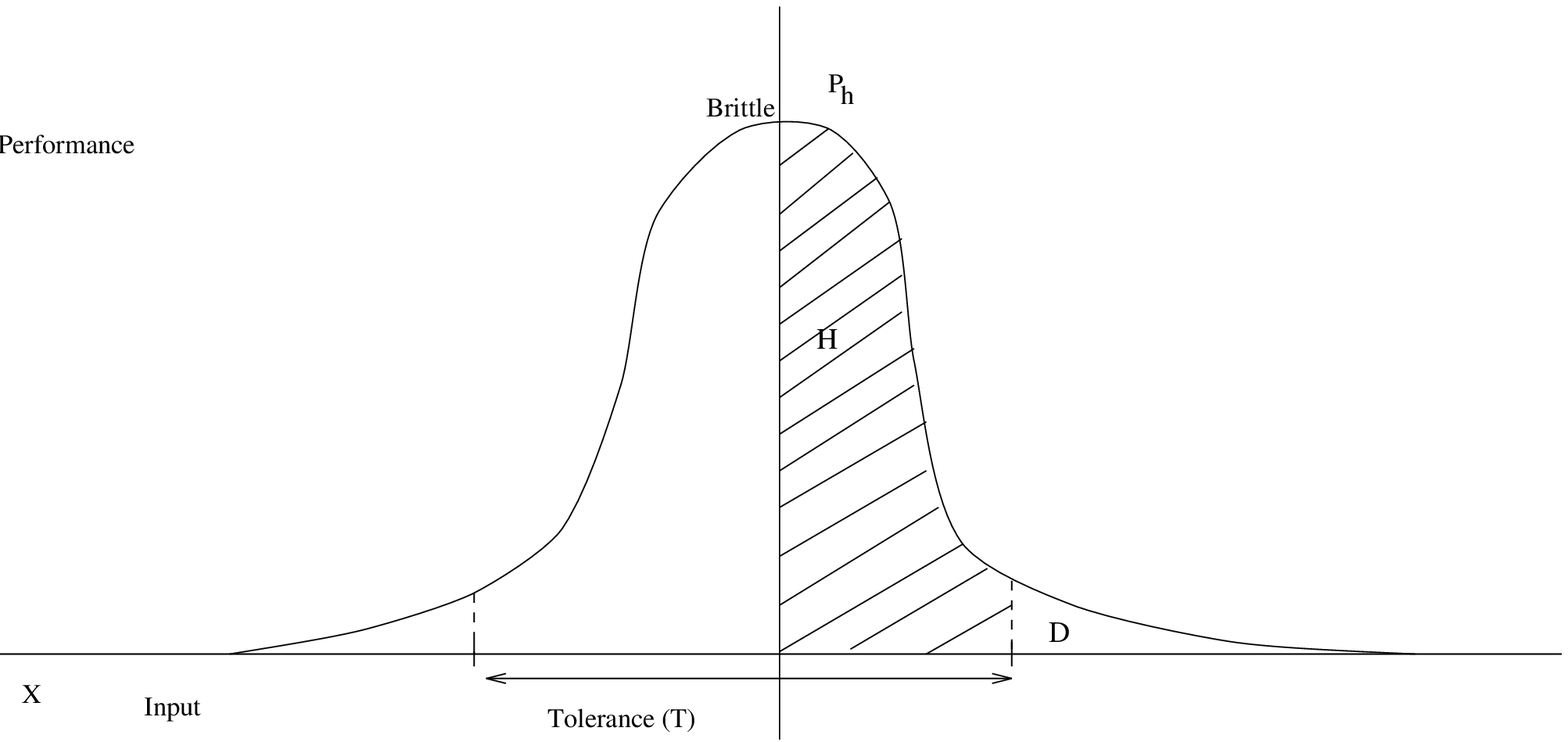,width=6.0in}}
        \caption{A Definition of System Brittleness.}
        \label{bdef}
\end{figure*}

Increasing both hardness and ductility increases the toughness of a 
material. Hardness is increased by deforming the crystal structure, either by
adding impurities to a homogeneous material or by rapid cooling of the
material after processing. In this work, increasing hardness of a
material is analogous to increasing the gain of the sub-components of a
system. Previous work has focused on the hardness of a system,
but relatively little on the ductility. For example, in choosing design 
parameters for a system \cite{Montgomery}, one examines the effect of high 
and low parameter values within the utility of normal operation of system 
performance ($U(normalOperation)$) and chooses those 
values which result in the best performance ($P_h$). However, the behavior and 
utility of the system when tolerance is exceeded ($U(robust)$) have rarely 
been examined. Certainly if time is considered,
then based on simple reliability theory the utility is shown in
Equation \ref{util} where $H$ and $D$ are shown in Figure \ref{bdef},
$x$ is a design parameter, $P[]$ is the probability of the event in the
brackets, and $U(normalOperation)$ is the utility to the user of the
system in normal operation, and $U(robust)$ is the utility to the user
of the system outside normal operation. As graceful degradation becomes a 
more desirable feature, the utility of area $D$ increases.
Let us define brittleness as the ratio of the hardness over the ductility 
which is the area H over D in Figure \ref{bdef}.

\begin{eqnarray}
Utility = P[x \in T] H U(normalOperation) + P[x \notin T] D U(robust) 
\label{util}
\end{eqnarray}

\begin{table*}[htbp]
\centering
\begin{tabular}{||l|l||}                                     \hline
\textbf{Materials Science}  & \textbf{Brittle Systems}    \\ \hline
\hline
Stress          &  amount parameter exceeds its tolerance \\ \hline
Toughness       &  system robustness \\ \hline
Hardness        &  level of performance within tolerance \\ \hline
Ductility       &  level of performance outside tolerance \\ \hline
Plastic Strain  &  system cannot recover from degradation \\ \hline
Reversible Strain & system can recover from degradation \\ \hline
Brittle Fracture & sudden steep decline in performance \\ \hline
Ductile Fracture & graceful degradation in performance \\ \hline
Brittleness     & ratio of hardness over ductility     \\ \hline
Deformation 	& degradation in performance		\\ \hline
Young's Modulus	& amount tolerance exceeded over degradation \\ \hline
\end{tabular}
\caption{\label{terms} Terminology.}
\end{table*}

At this point it must be mentioned that
some causes of brittle fracture may be more difficult to deal with than
others. For example the sudden loss of performance can be due to a 
catastrophe \cite{Saunders,IB-D84160}. Catastrophe Theory is
essentially the study of singularities; in this work it would be
one of many causes for brittle behavior. The connection between
Catastrophe Theory and Brittle Systems is only one of the many
areas that need to be explored in this new research area.
\section{Sensitivity}

\begin{figure*}[htbp]
        \centerline{\psfig{file=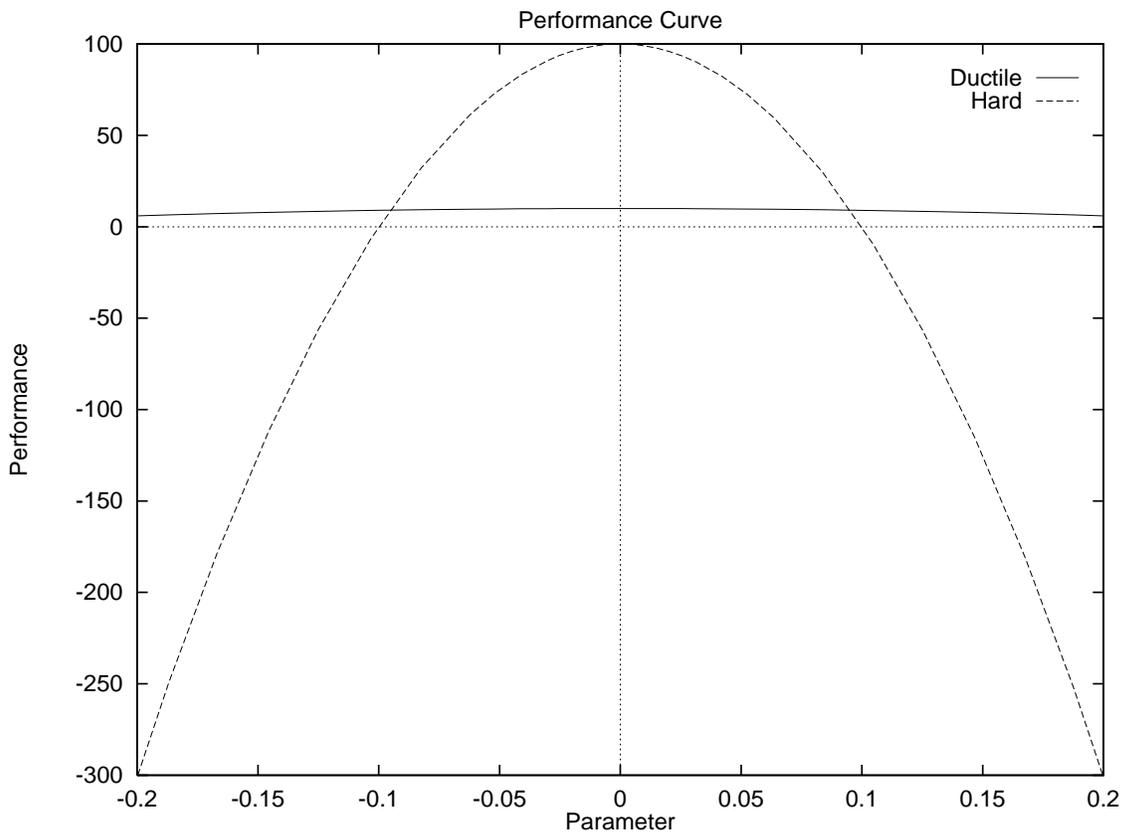,width=6.0in}}
        \caption{An Example System.}
        \label{esys}
\end{figure*}

\begin{figure*}[htbp]
        \centerline{\psfig{file=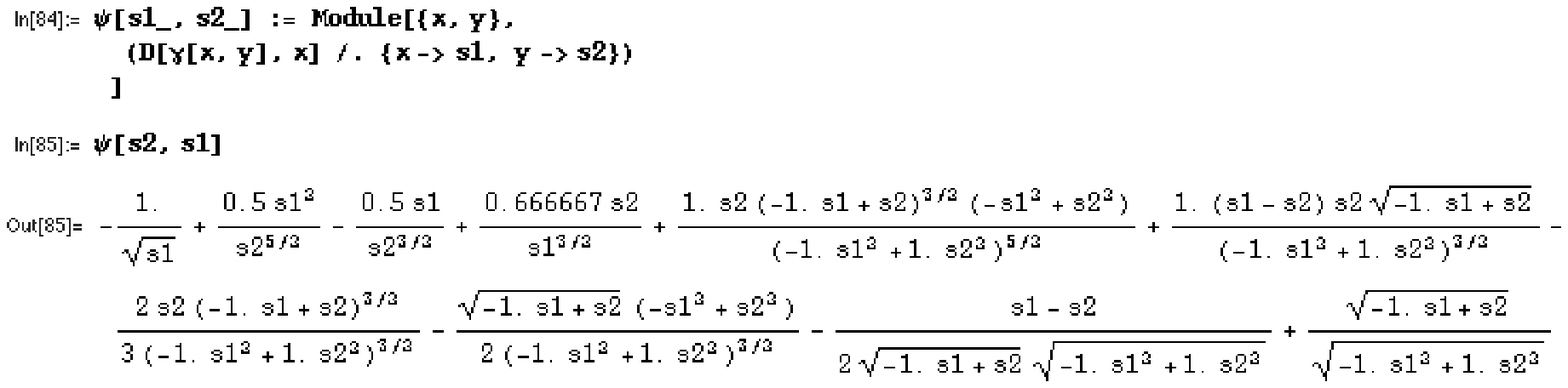,width=6.0in}}
        \caption{Ductility Sensitivity Definition.}
        \label{ddef}
\end{figure*}

As a first step, design parameters, $X_p$, which affect ductility
must be identified.
The sensitivity of ductility to a particular parameter is characterized
by $\psi$, as shown in Equation \ref{pdeff} and \ref{peff}.
In Equation \ref{pdeff}, $\gamma$ is the shaded area $A-B$ in Figure 
\ref{bintro}, which is a function of two values, $x_1$ and $x_2$, of 
a single design parameter, $X_p$.
In Equation \ref{peff}, the sensitivity of ductility is defined as the rate of
change of the difference of $A-B$. Figure \ref{esys} shows two
curves for the same system, one curve which is brittle, the other
robust. A function which returns the value of the ductile sensitivity
is implemented in Mathematica \cite{Wolfram:MSD91} in Figure \ref{ddef}. 
In Figure \ref{ddef}, $\psi$ takes two arguments, $s1$ and $s2$ which 
are two values of a single design parameter, $x_1$ and $x_2$. The 
Mathematica module returns the partial 
derivative of $\gamma$ as shown in the bottom of Figure \ref{ddef}.

\begin{figure*}[htbp]
        \centerline{\psfig{file=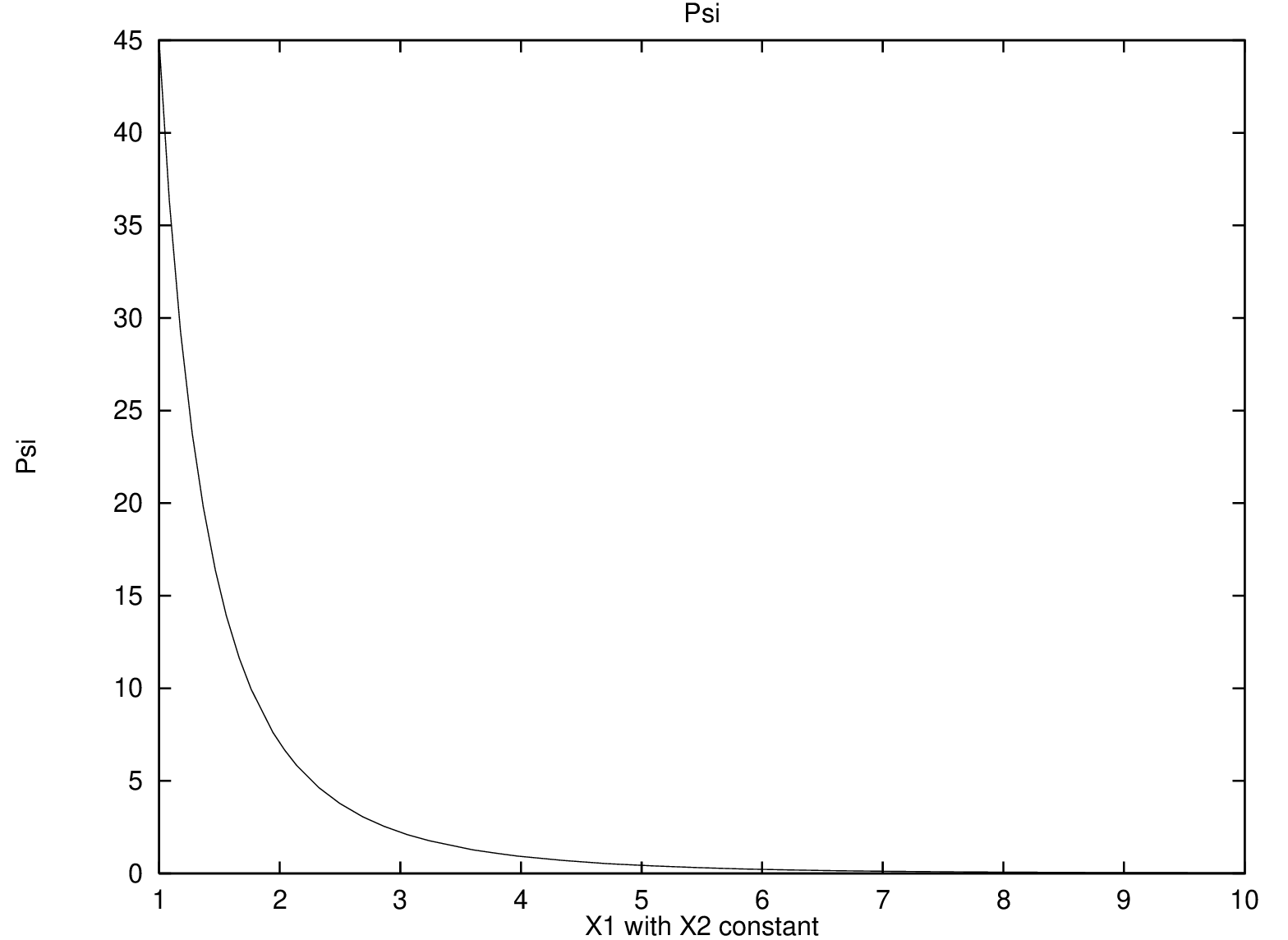,width=4.0in}}
        \caption{Ductile Sensitivity.}
        \label{duc}
\end{figure*}

In Figure \ref{duc}, the value of the ductility sensitivity is
shown for the system from Figure \ref{esys} as a function of the 
difference between $x_1$ and $x_2$. In Figure \ref{duc}, $x_2$ is 
constant and $x_1$ varies. As $x_1$ and $x_2$ become equal, $\Psi$ 
goes to zero. This is because the performance curves become the
same and the area $A-B$ disappears. Also, when the values of
$x_1$ and $x_2$ are far apart, the area $A-B$ becomes large and
the rate of change of the area becomes large. Note that because
of the implementation of the Mathematica module which computes
$\gamma$, the order of the arguments to the Mathematica function in 
Figure \ref{ddef} is significant.

\begin{equation}
\label{pdeff}
\gamma(x_1,x_2) = {A - B}
\end{equation}

\begin{eqnarray}
\psi(x_1, x_2) = {{\partial \gamma(x_1,x_2)} \over {\partial x_1}} & & 
\label{peff}
\end{eqnarray}

\section{System Energy}

\begin{figure*}[htbp]
        \centerline{\psfig{file=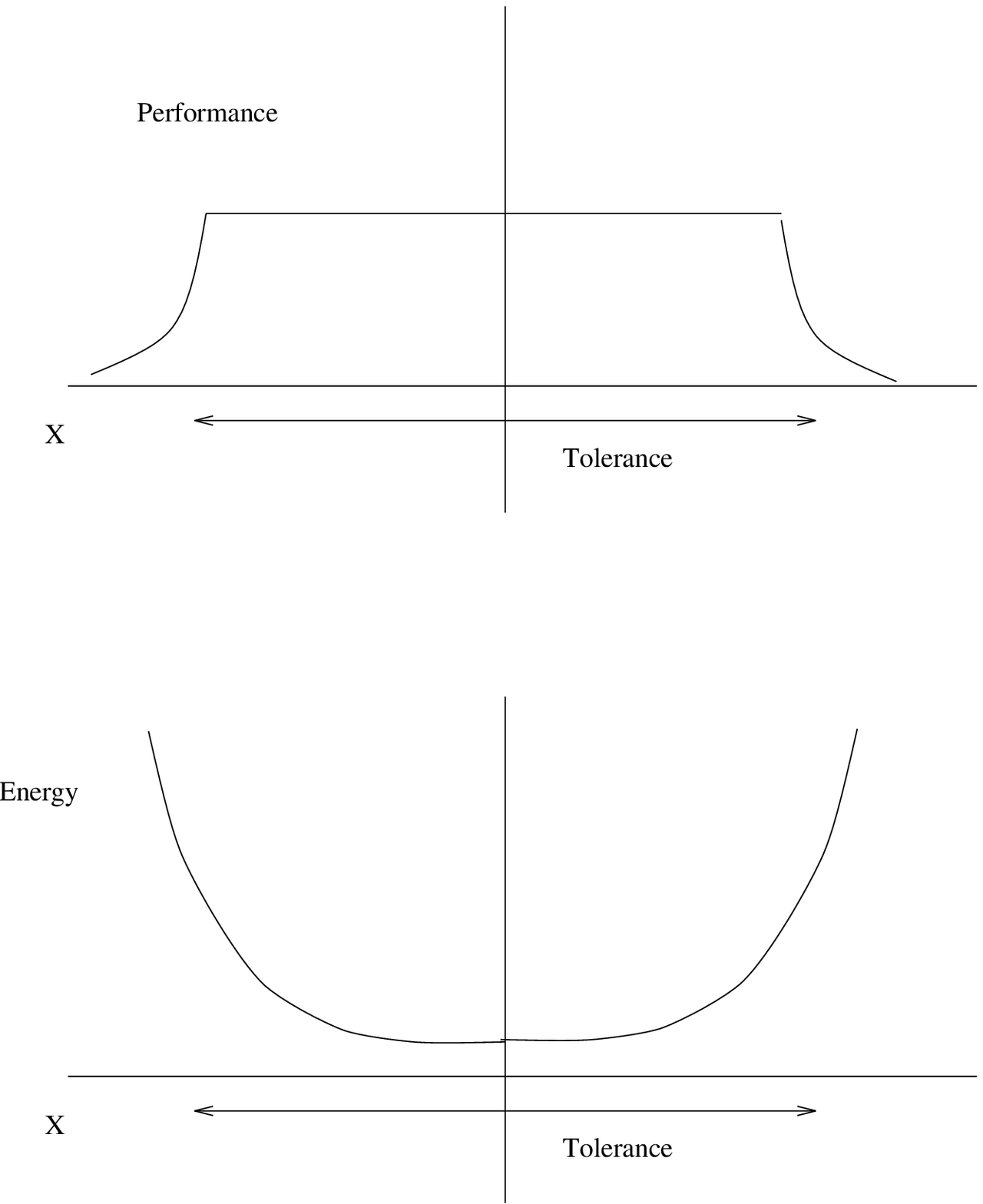,width=4.0in}}
        \caption{System Performance and Energy.}
        \label{energy}
\end{figure*}

\begin{figure*}[htbp]
        \centerline{\psfig{file=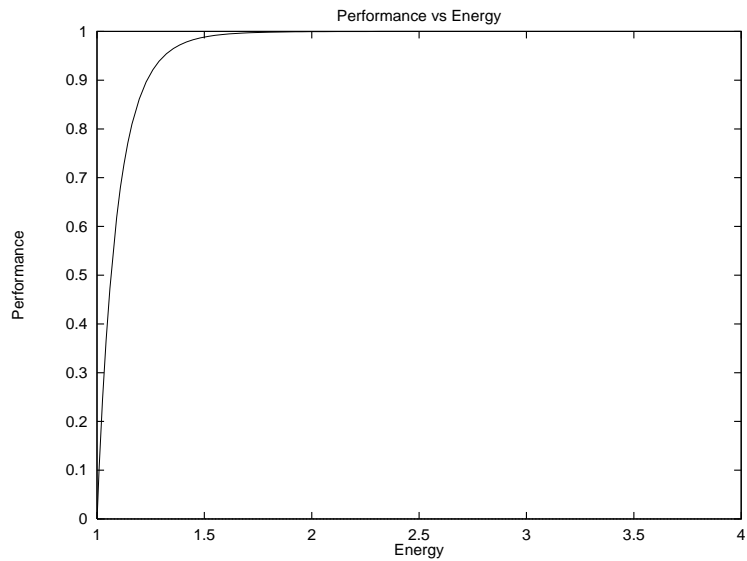,width=4.0in}}
        \caption{Performance and Energy.}
        \label{epgraph}
\end{figure*}

As a digital system approaches the edges of it operating tolerance the
energy required to maintain the performance increases as shown in Figure
\ref{energy}. Consider quality of service on a router in a communications 
network. The energy required to forward a packet is routinely modeled as
directly proportional to the length of the packet. 
As the load increases, input queues begin to fill to capacity
and packets are dropped because computational energy is not sufficient
to keep up with the load. In this case, performance is the probability of
not dropping a packet and energy is the processing power which is 
directly proportional to the packet service rate, $\mu$. In an 
M/M/1 queue, a direct relation between performance and
energy is shown in Equation \ref{perfprob} where $n$ is the expected queue
size, $N$ is the maximum queue capacity and
$\rho = {\lambda \over \mu}$. The result is graphed in Figure
\ref{epgraph}.

\begin{equation}
P[n<N] = \sum_0^{N-1} \rho^n(1-\rho)
\label{perfprob}
\end{equation}

\section{Brittle Sub-Components}

\begin{figure*}[htbp] 
        \centerline{\psfig{file=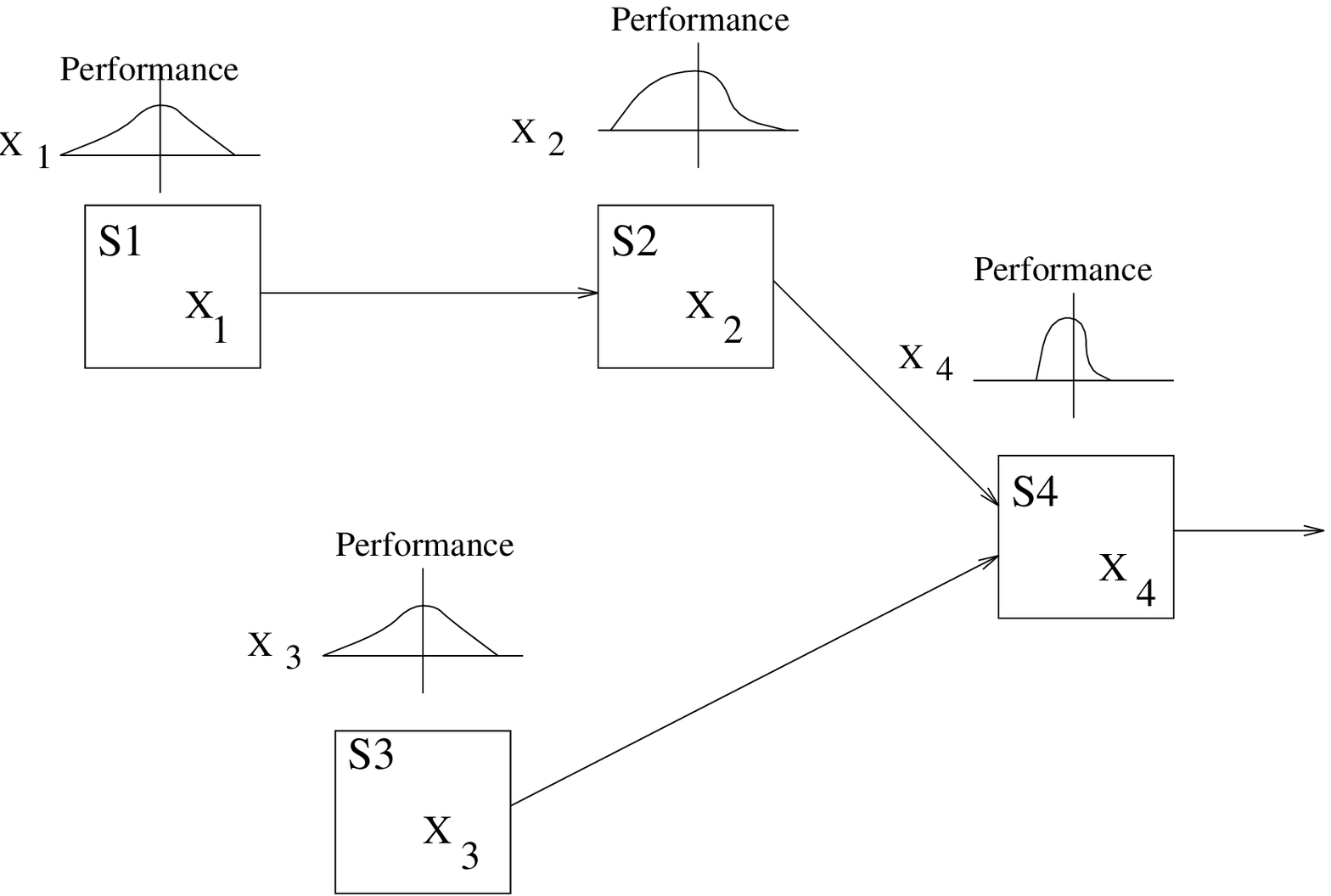,width=4.0in}} 
        \caption{Brittle Subsystem Components.}
        \label{bnet} 
\end{figure*}

Consider a system whose sub-components exhibit various degrees of
ductility as defined above. Just as adding impurities to a pure metal causes
it to become stronger but more brittle, the addition of more efficient
but also more sensitive components to a system causes the system to
increase performance within its operating range, but become less
ductile. How do the effects of ductility propagate among the
sub-components to influence the ductility of the entire system? Assume
the performance response curve is known for each sub-component and that
the output from one component feeds into the input of the next component
as shown in Figure \ref{bnet}. Assume that the sub-component output
performance cannot be better than any of its inputs. Then the performance 
curve for the output of each sub-component is the minimum  of the input 
sub-component performance curve and the current component performance curve.

\begin{figure*}[htbp]
        \centerline{\psfig{file=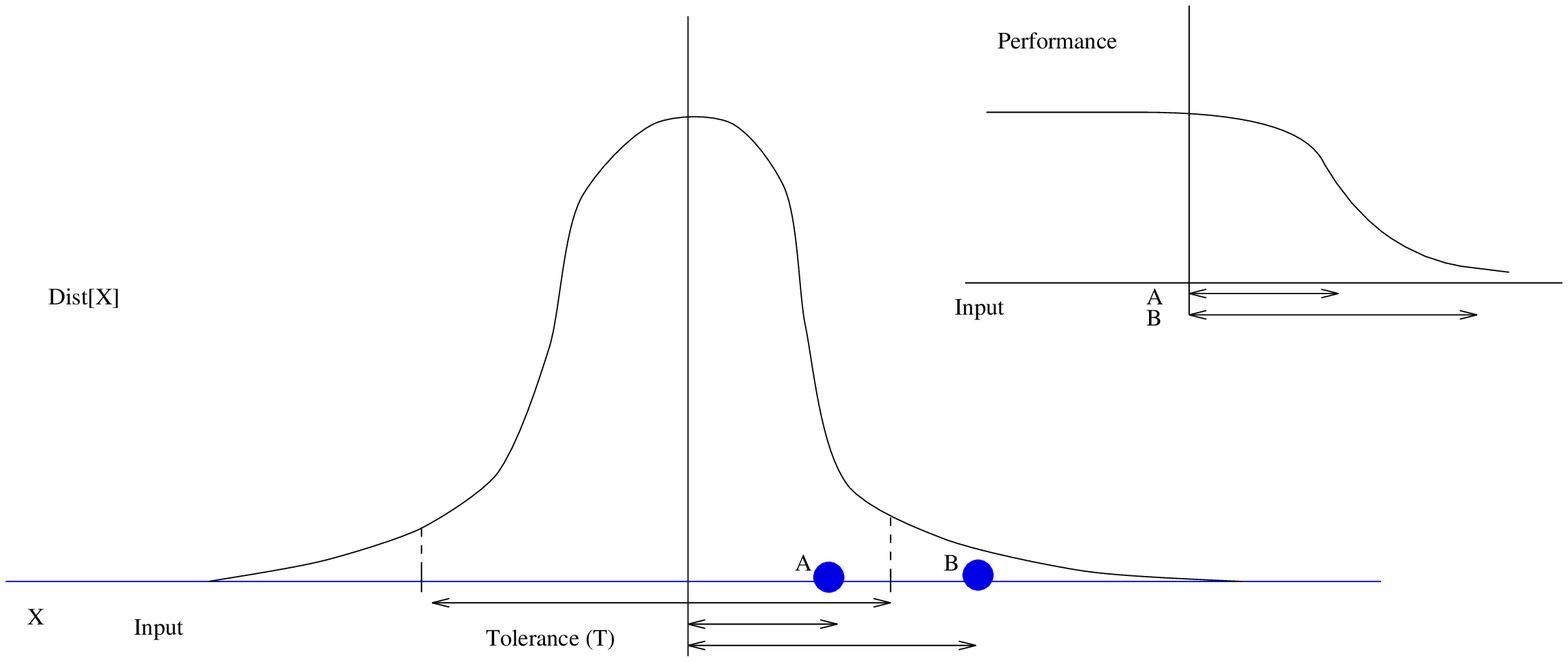,width=6.0in}}
        \caption{Distribution and Performance.}
        \label{poutdiagram}
\end{figure*}

\begin{figure*}[htbp]  
        \centerline{\psfig{file=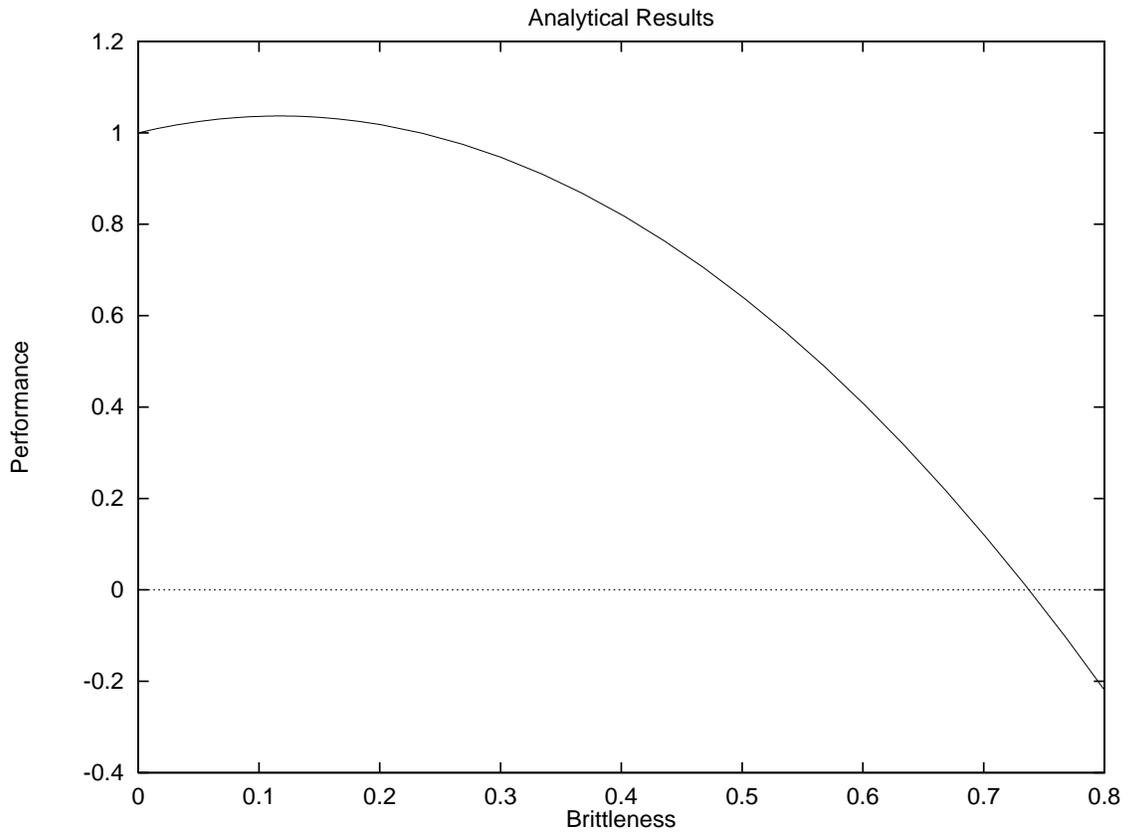,width=6.0in}}  
        \caption{Analytical Brittle Subcomponent Result.}
        \label{analpcurve}  
\end{figure*}

The hardness component of the
brittleness enhances the performance when values are within tolerance 
and low ductility degrades the performance when values are
out of tolerance. The amount of degradation depends on the amount by
which the tolerance was exceeded. This is illustrated in Figure
\ref{poutdiagram} and is stated in Equation \ref{pout},
where $b$ is the brittleness, $P_{in}$ is the input performance, $T$ is
the set of in-tolerance values, $x$ is a state parameter, $E[]$ is the
expected value, and $P_{out}$ is the output performance. The result of 
Equation \ref{pout} is plotted in Figure \ref{analpcurve}.

\begin{equation}
\label{pout}
P_{out} = (P_{in}+P_{in} b)Prob[x \in T] + (P_{in} - P_{in} b 
E[x-max[T]])Prob[x \notin T]
\end{equation}

As $b$ decreases in a non-brittle system, $T$ increases. It is this
relationship between $b$ and $T$ which is the principal focus of brittle
systems analysis. Assume the simple case of a normally distributed
performance distribution, then Equation \ref{normpout} shows how
Equation \ref{pout} can be refined. $N(\eta,\sigma)$ is a normal
distribution with an average of $\eta$ and variance of $\sigma$ and
$R.V._{N(\eta,\sigma)}$ is a random variable with distribution
$N(\eta,\sigma)$.

\begin{equation}
\label{normpout}
P_{out} = (P_in +P_{in} b) Prob[R.V._{N(\eta,\sigma)} \le T] + (P_in - b \eta)
1.0 - Prob[R.V._{N(\eta,\sigma)} \le T]
\end{equation}

\begin{figure*}[htbp]
        \centerline{\psfig{file=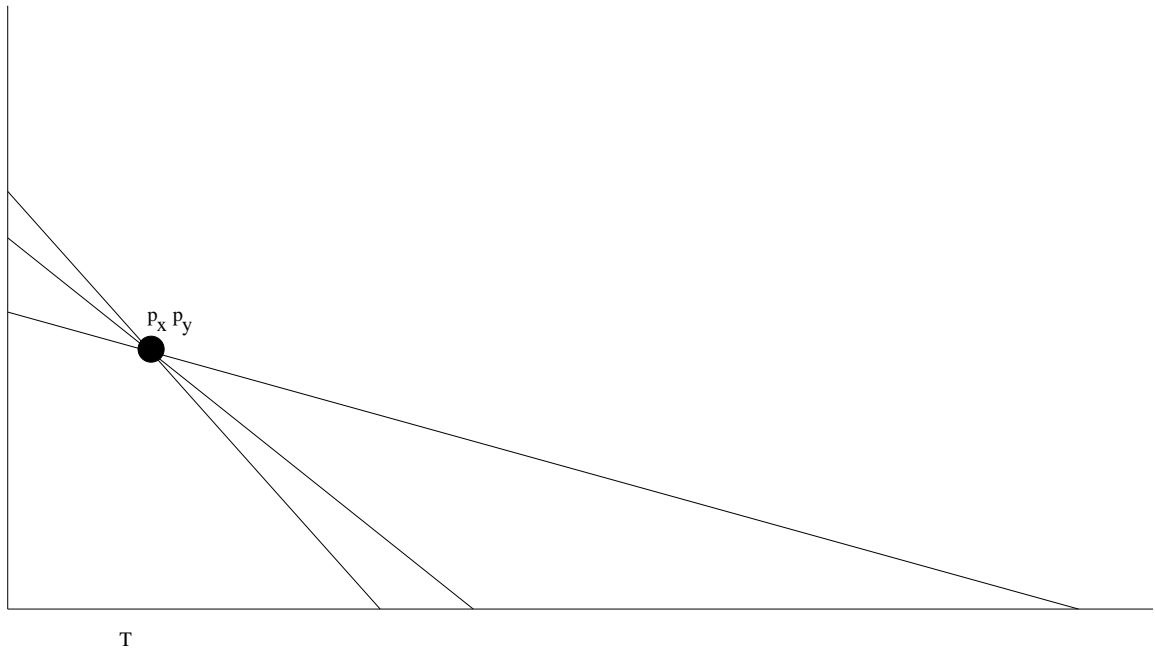,width=6.0in}}
        \caption{Relationship Between $b$ and $T$.}
        \label{linbfig}
\end{figure*}

As $b$ decreases we assume that the system is non-brittle
so that $T$ increases. Assume that $b$ is linear, then $T$ increases as
shown in Figure \ref{linbfig} and Equation \ref{linbeq}.

\begin{equation}         
\label{linbeq} 
T = p_x + {1 \over b} p_y
\end{equation}

\begin{figure*}[htbp]
        \centerline{\psfig{file=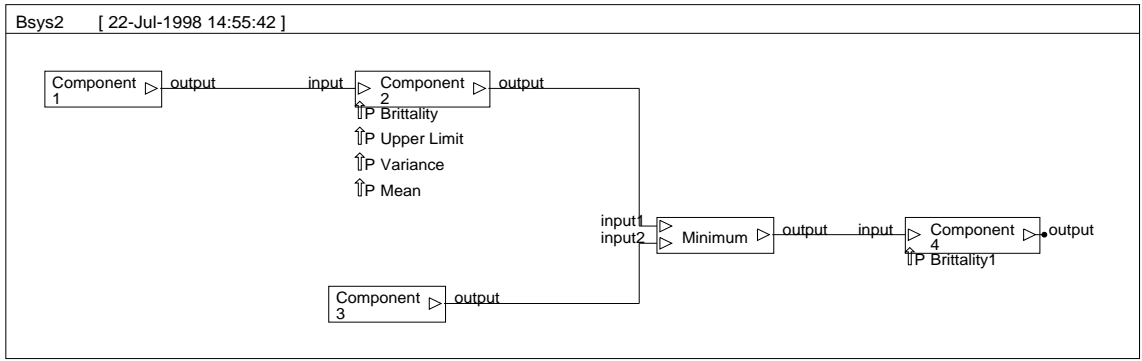,width=6.0in}}
        \caption{BONeS Model.}
        \label{bmodel}
\end{figure*}

A BONeS model has been developed to examine brittle sub-components as
shown in Figure \ref{bmodel} which models Figure \ref{bnet}. A 
BONeS data structure contains the performance or
quality of the input to a component. A normal random number generator
produces a value with a specified mean and variance, in this case 10.0
and 3.0 respectively. The difference between the random number and the
upper limit (11.0) is computed. If the normal random number is greater
than the upper limit, then the performance value of the input data 
structure is reduced by the brittleness multiplied by the amount by which the
tolerance was exceeded. If the normal random number is within
tolerance  then the input data structure is increased by an amount
proportional to the brittleness. 

\begin{figure*}[htbp]  
        \centerline{\psfig{file=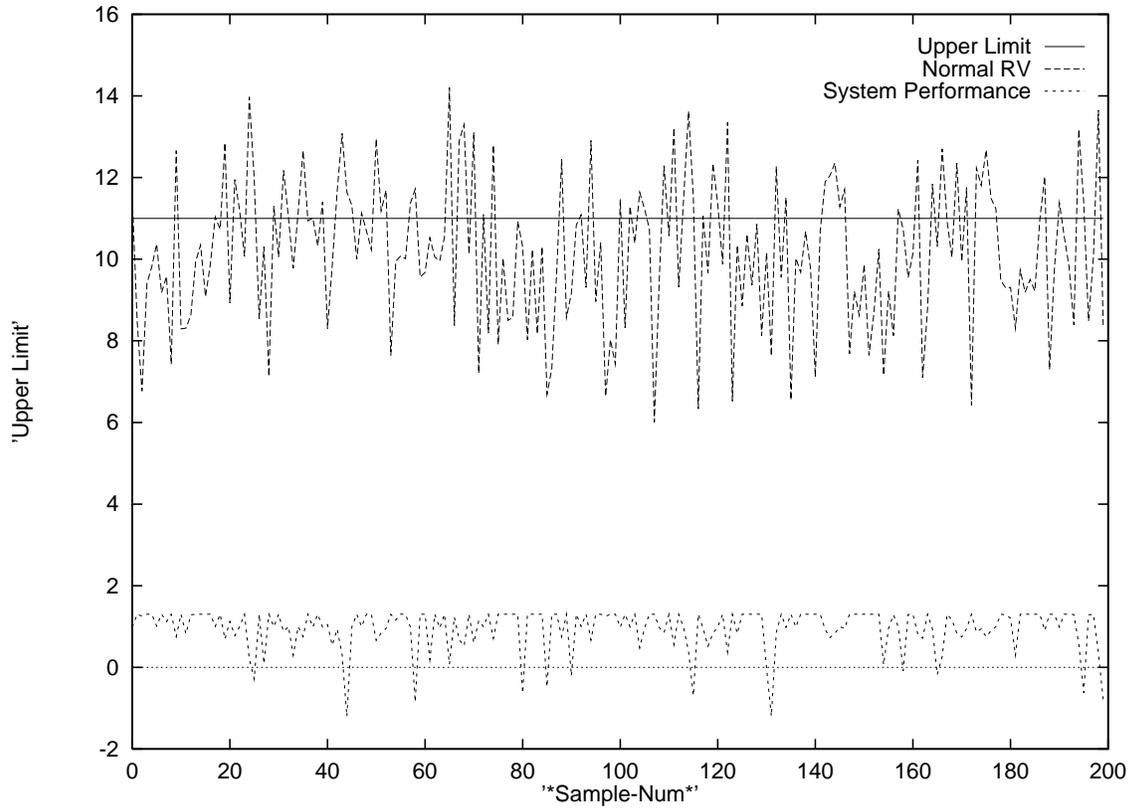,width=6.0in}}
        \caption{System Performance.}
        \label{sysperf}  
\end{figure*}

\begin{figure*}[htbp]  
        \centerline{\psfig{file=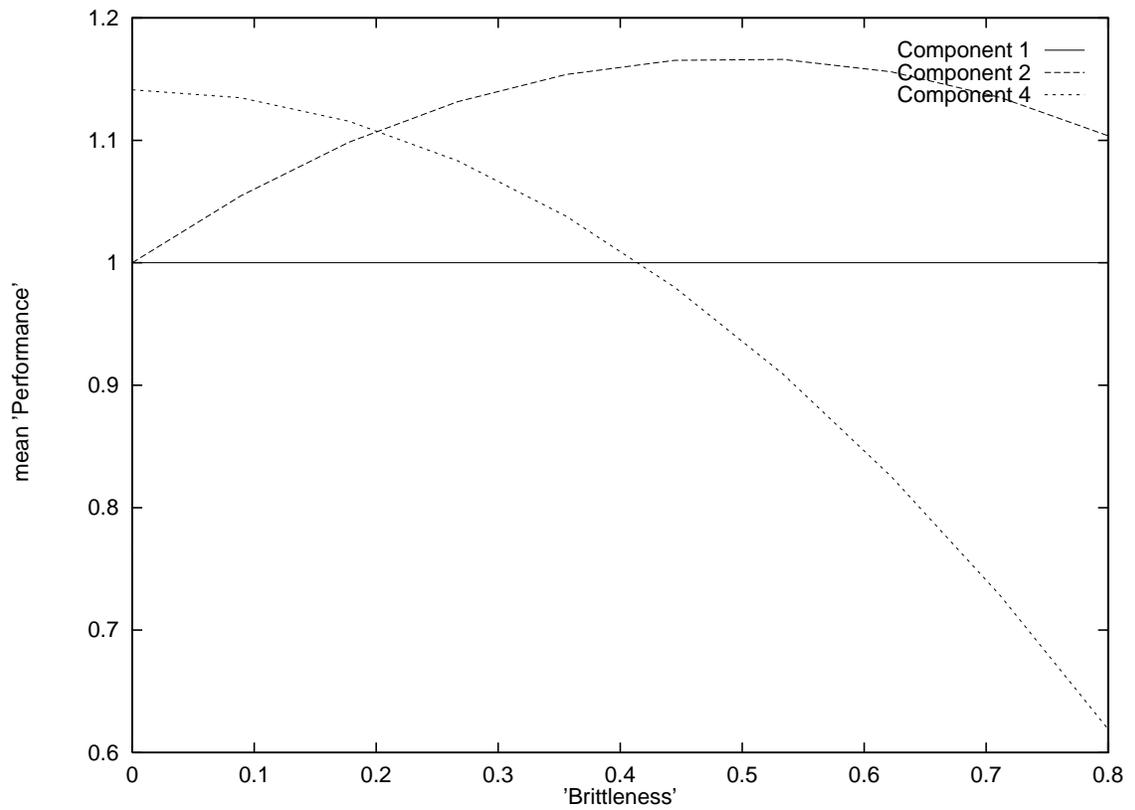,width=6.0in}}
        \caption{Performance Curve.}
        \label{pcurve}  
\end{figure*}

The results are shown in Figure \ref{sysperf} for the normal random
number values, the upper limit, and the system performance as a function 
of the consecutive order in which each of the values were sampled. The 
brittleness is varied from zero to
0.8 and the results are averaged. Clearly the performance degrades when
the normal values exceed the upper limit. Figure \ref{pcurve} shows the
performance results for the sub-components and the entire system from
Figure \ref{bmodel}. Components 1 and 3
generate data structures with a performance value of one. An
intermediate component, Component 2, has a brittleness which varies 
from zero to 0.8.
The final output component, Component 4, has a brittleness of 0.3. The
analytical results from Figure \ref{analpcurve} and the simulated
system performance curve from Figure \ref{pcurve} are in close
agreement. Although Component 2 performance
improves when the brittleness is between 0.2 and 0.5, Component 4, which
is the system performance, declines. This is because Component 4
performance depends on the minimum performance input which comes from 
Component 3, an initial input component that always generates a
performance of one.

\begin{figure*}[htbp]
        \centerline{\psfig{file=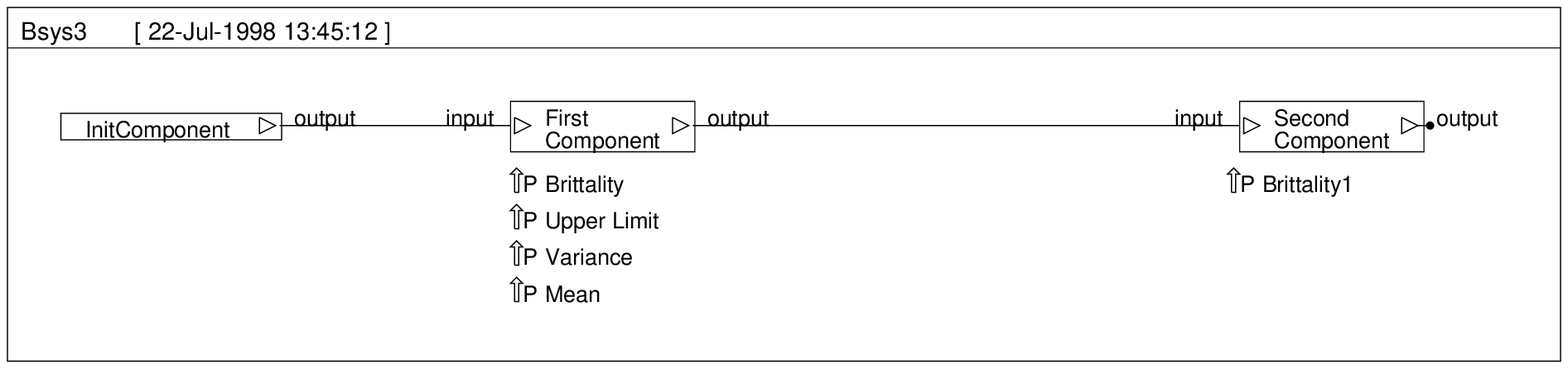,width=6.0in}}
        \caption{Brittleness Location Experiment.}
        \label{dloc}
\end{figure*}

\begin{figure*}[htbp]
        \centerline{\psfig{file=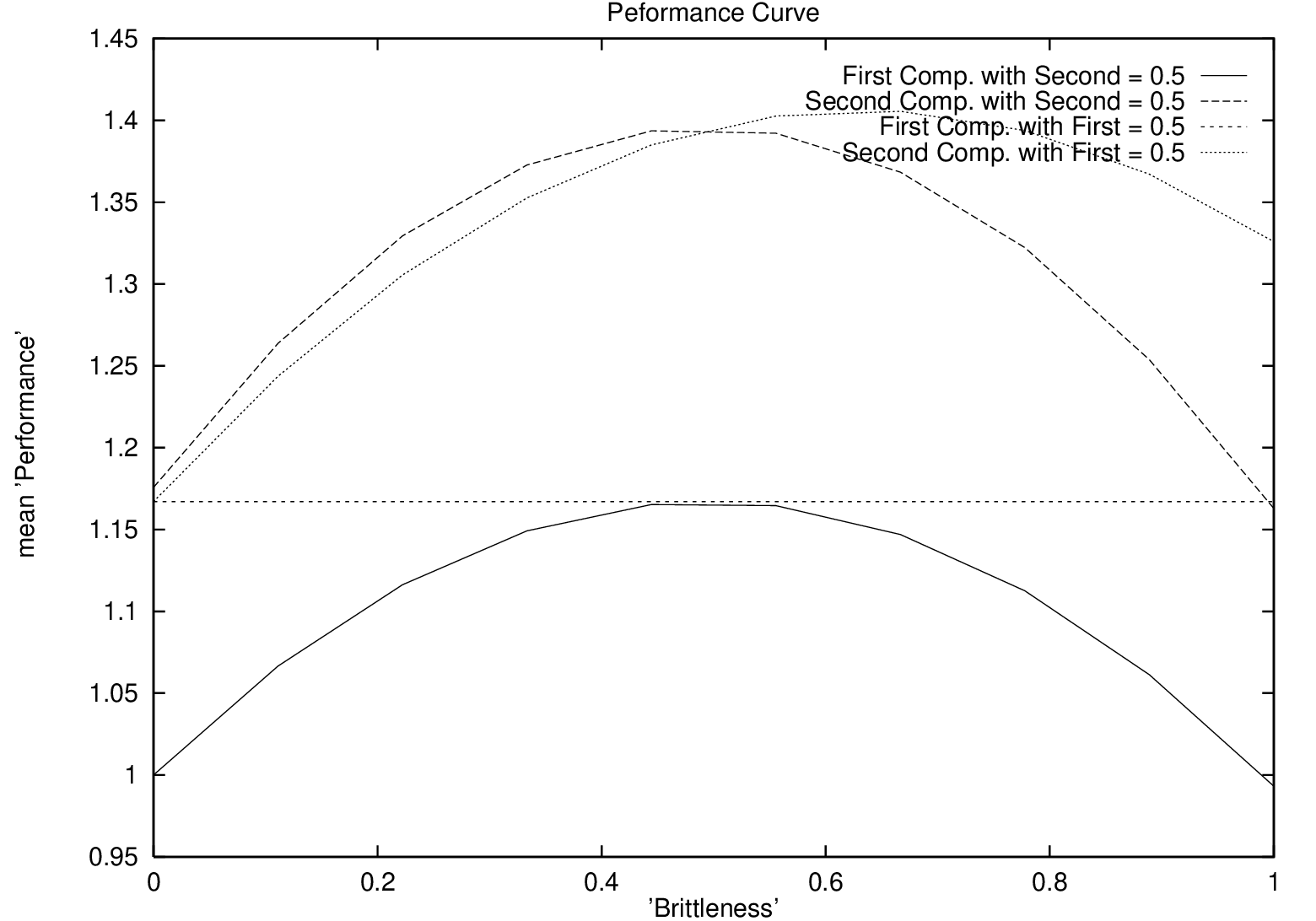,width=6.0in}}
        \caption{Brittleness Location Results.}
        \label{fvss}
\end{figure*}

If the ductility of sub-components can be controlled, how should the
brittleness be adjusted among the sub-components? One line of reasoning
yields the result that in systems run near the maximum operating tolerance, 
better performance will be achieved with highly brittle components placed 
near the outputs of the system. This is because there is then less 
chance for the highly brittle components to effect the other sub-components. 
The next simulation, shown in Figure \ref{dloc} examines this question.
The brittleness of the first component is varied
from zero to one and the second component remains at a brittleness of
0.5. The results are shown in Figure \ref{fvss}. The results are also 
shown in the same figure for the first component brittleness of 0.5 and the 
second component brittleness varying from zero to one. Figure \ref{fvss}
indicates that the best performance curve results when the more highly 
brittle component is the last component in the chain.

\section{An Example of Ductility in a Communications Network}
\label{example}

The following applications which exhibit brittle behavior have
been chosen as simple examples so that the ideas presented in this work,
rather than the details of the applications, can be investigated. These
examples will be examined in more detail as this work progresses.

\subsection{Adaptive Multimedia}

\begin{figure*}[htbp]
        \centerline{\psfig{file=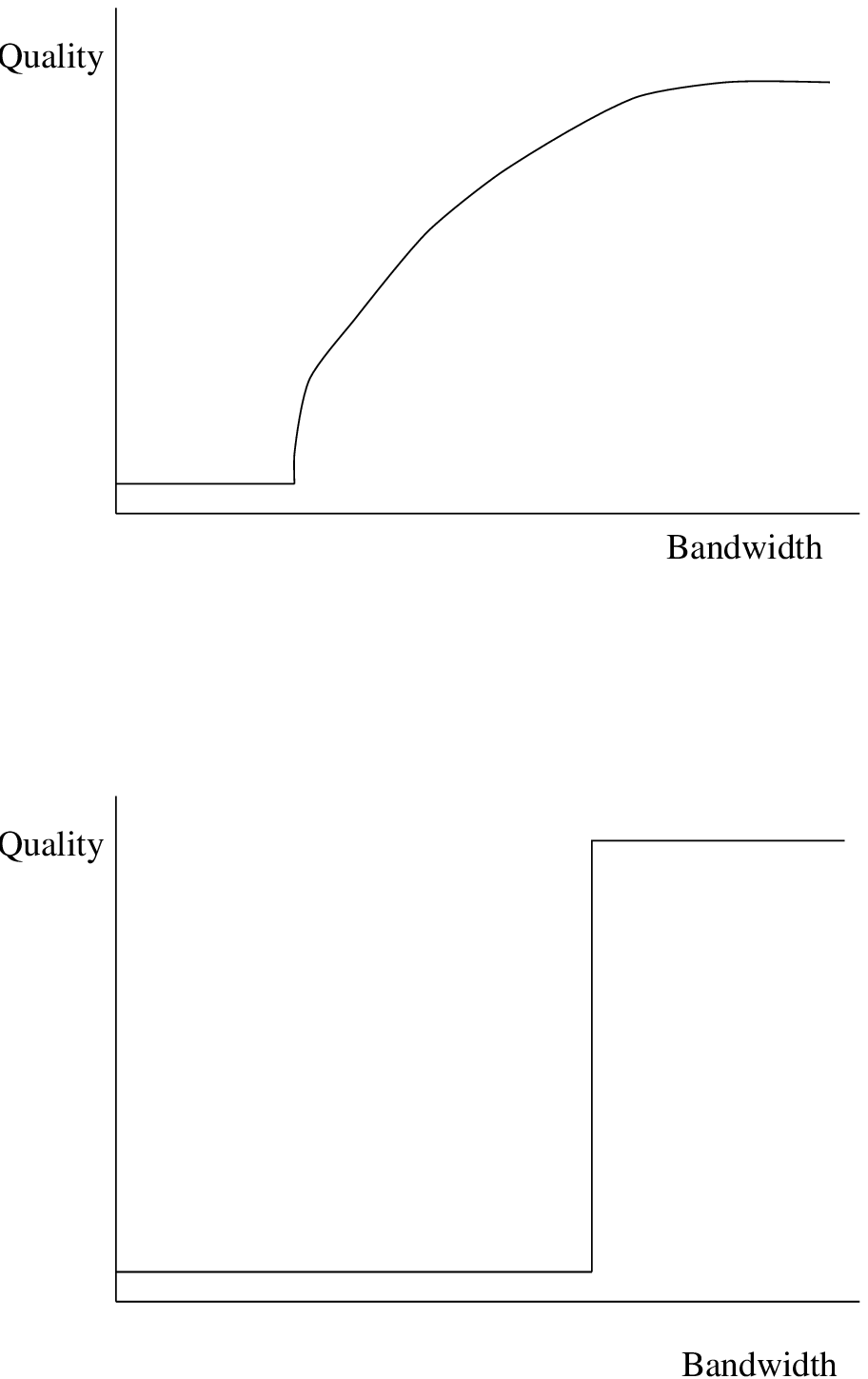,width=2.5in}}
        \caption{Ductile Network Applications (Top Graph).}
        \label{qbplot}
\end{figure*}

Current network applications, especially multimedia applications, have
performance which degrades rapidly after bandwidth is reduced beyond a 
certain point. In \cite{Lee} it is suggested that if applications can be 
developed which degrade gracefully with respect to loss in bandwidth as
shown in Figure \ref{qbplot},
then the network can be designed to maintain bandwidth within the required
bounds on a best effort basis.
A solution recommended in \cite{Lee} is for the network
to keep a certain amount of bandwidth in reserve. However,
the more bandwidth kept in reserve, the less that remains to support
the network as a whole. Thus the amount of reserve bandwidth is
the greatest factor affecting ductility in this example. As the
value of reserve bandwidth increases, the number of 
users which can be supported is reduced, but fewer calls in progress are
disconnected.

\subsection{Packet Recovery: Stop And Wait System}

\begin{figure*}[htbp]
        \centerline{\psfig{file=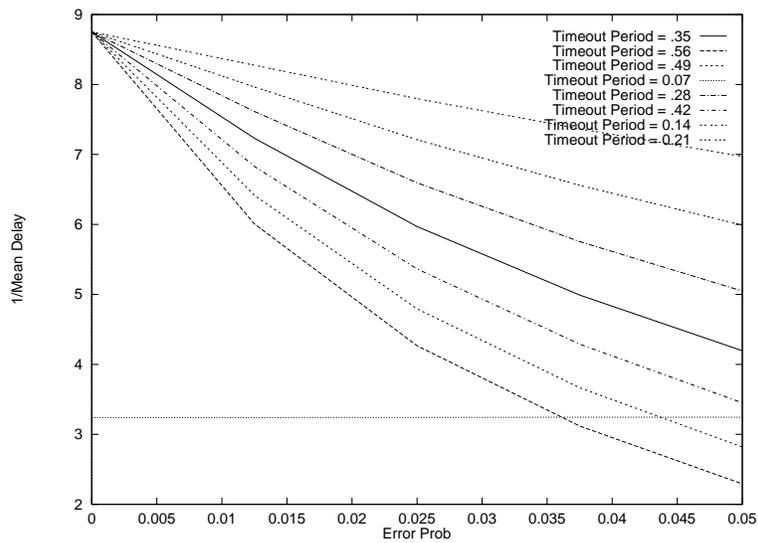,width=4.0in}}
        \caption{Delay, Probability of Error, and Timeout in a Stop And Wait System.}
        \label{saw}
\end{figure*}

The second example is recovery from packet loss in an Automatic Repeat
Request (ARQ) link shown in Figure \ref{saw}.
We consider two types of packets: packets with a large delay and
packets which are lost. Setting a high time-out value results in
better performance for packets which have a high ratio of delay
to loss, but degrades rapidly as the ratio approaches zero.

\subsection{TDMA Reservation System}

\begin{figure*}[htbp] 
        \centerline{\psfig{file=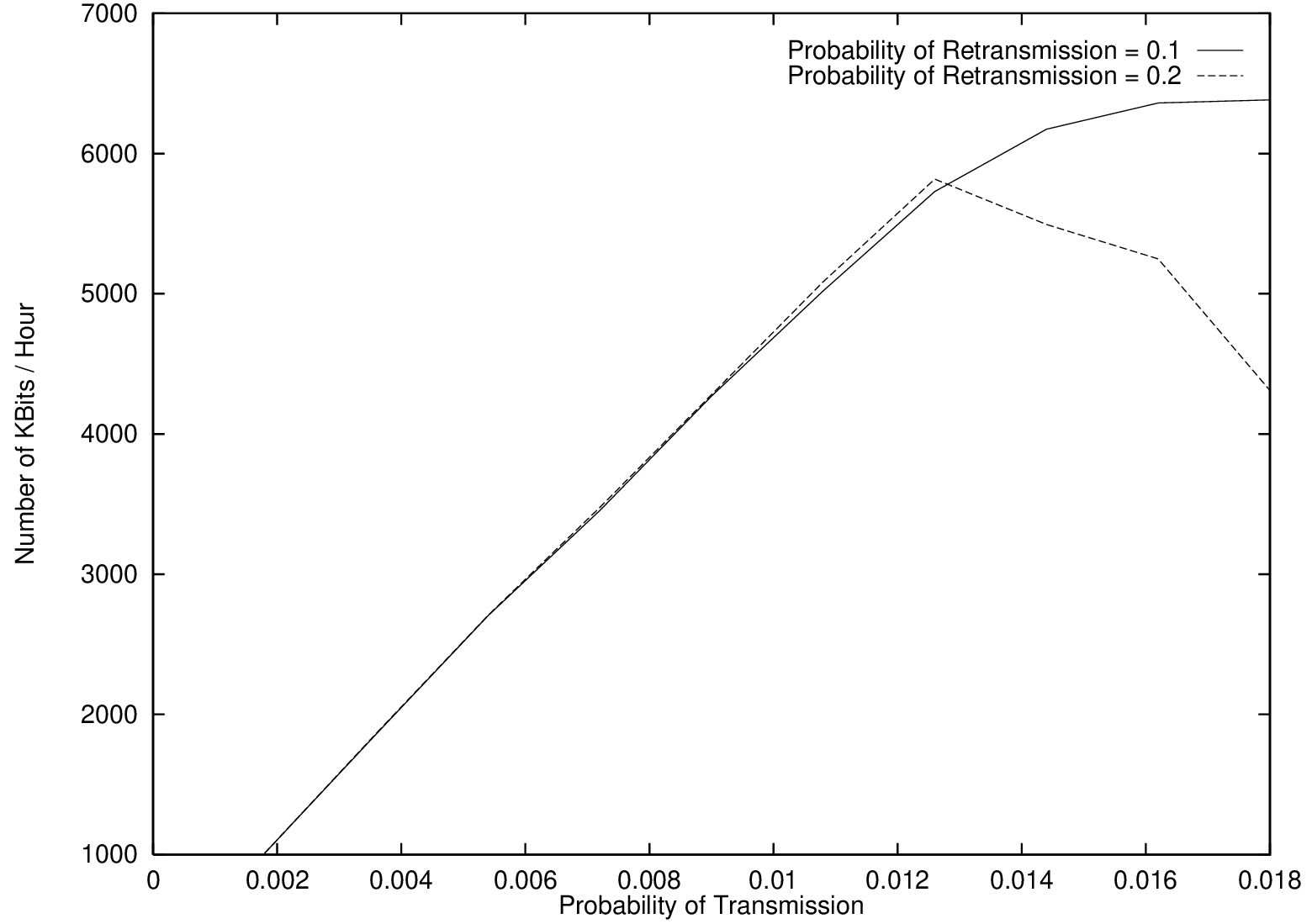,width=4.0in}}
        \caption{TDMA Probability of Transmission and Transmission Rate.}
        \label{tdma}
\end{figure*}

Figure \ref{tdma} shows transmission rate versus probability of transmission 
for two values of retransmission. The lower valued setting for retransmission
has higher performance, however, the higher valued retransmission
setting is slightly more robust around a probability of 0.013.

\subsection{Mobile Cellular Telephone System}

\begin{figure*}[htbp]
        \centerline{\psfig{file=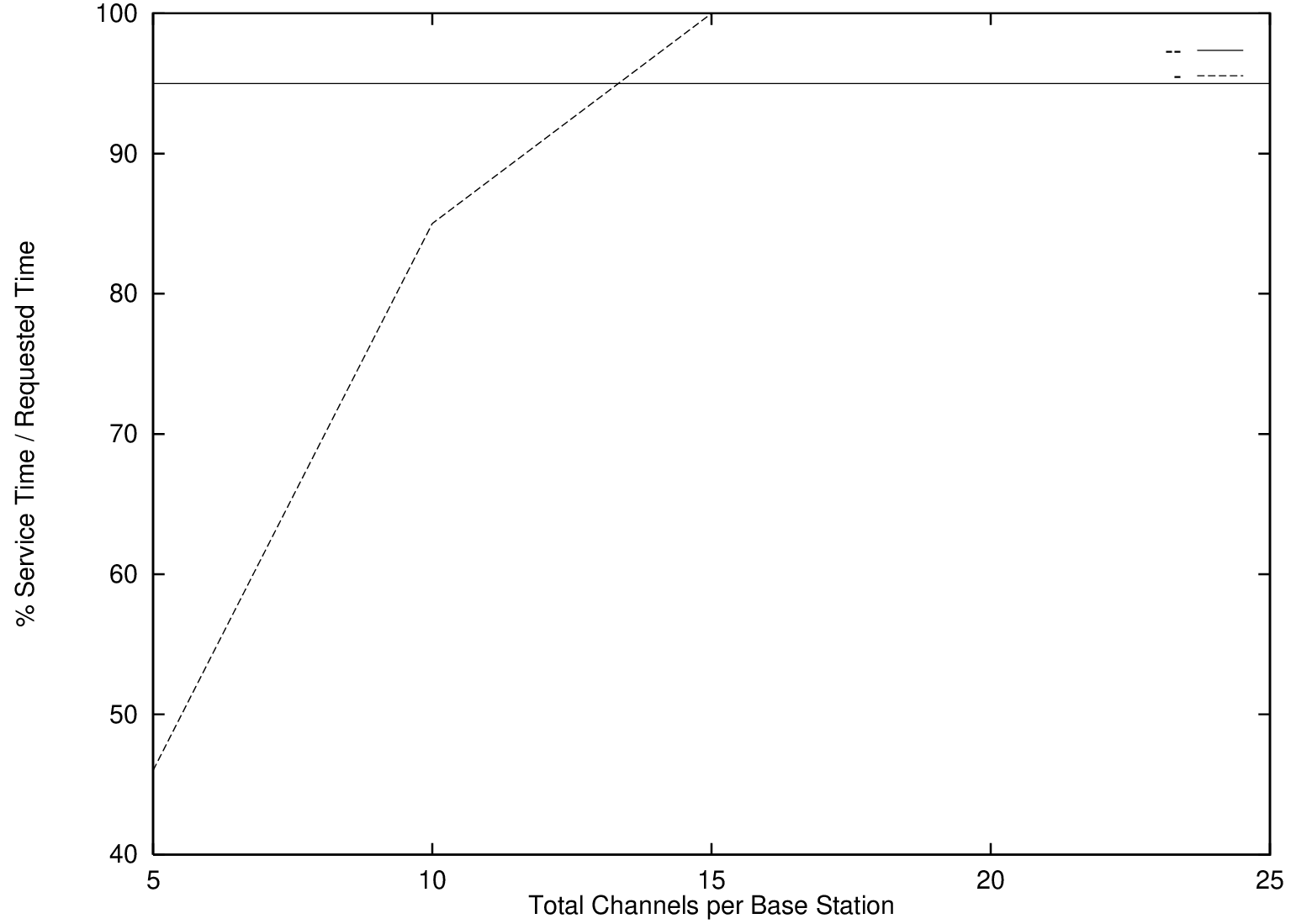,width=4.0in}}
        \caption{Grade of Service versus number of Channels per Base
Station.}
        \label{mobile}
\end{figure*}

Figure \ref{mobile} shows grade of service versus channels per base station.

\subsection{Buffer Capacity}

Another example of a brittle system involves choosing buffer
capacity in a data communications system.

\subsection{Backlogged Packets in Slotted ALOHA}

In Slotted ALOHA, data packet transmission occurs using equal sized
packets within equally divided time slots. If two or more users transmit
within a given time slot, a collision occurs; the packet will be
retransmitted in a following time slot with a given probability. 
This example of a brittle system exhibits catastrophic behavior
\cite{Nelson:1987:SCT}. Let $p_0$ be the probability that a packet to
be transmitted finds an empty cell, and $p_1$ be the probability that
after a collision, the cell attempts retransmission. The design
parameters are $p_0$ and $p_1$ and the number of packets waiting for 
retransmission is the state. A graph of $p_0$ and $p_1$ 
forms a cusp and all the classic symptoms of catastrophe are present,
namely, bifurcation, sudden jumps, hysteresis, inaccessibility, and
divergence.

\subsection{Variable Window Flow Control}

\begin{figure*}[htbp]
        \centerline{\psfig{file=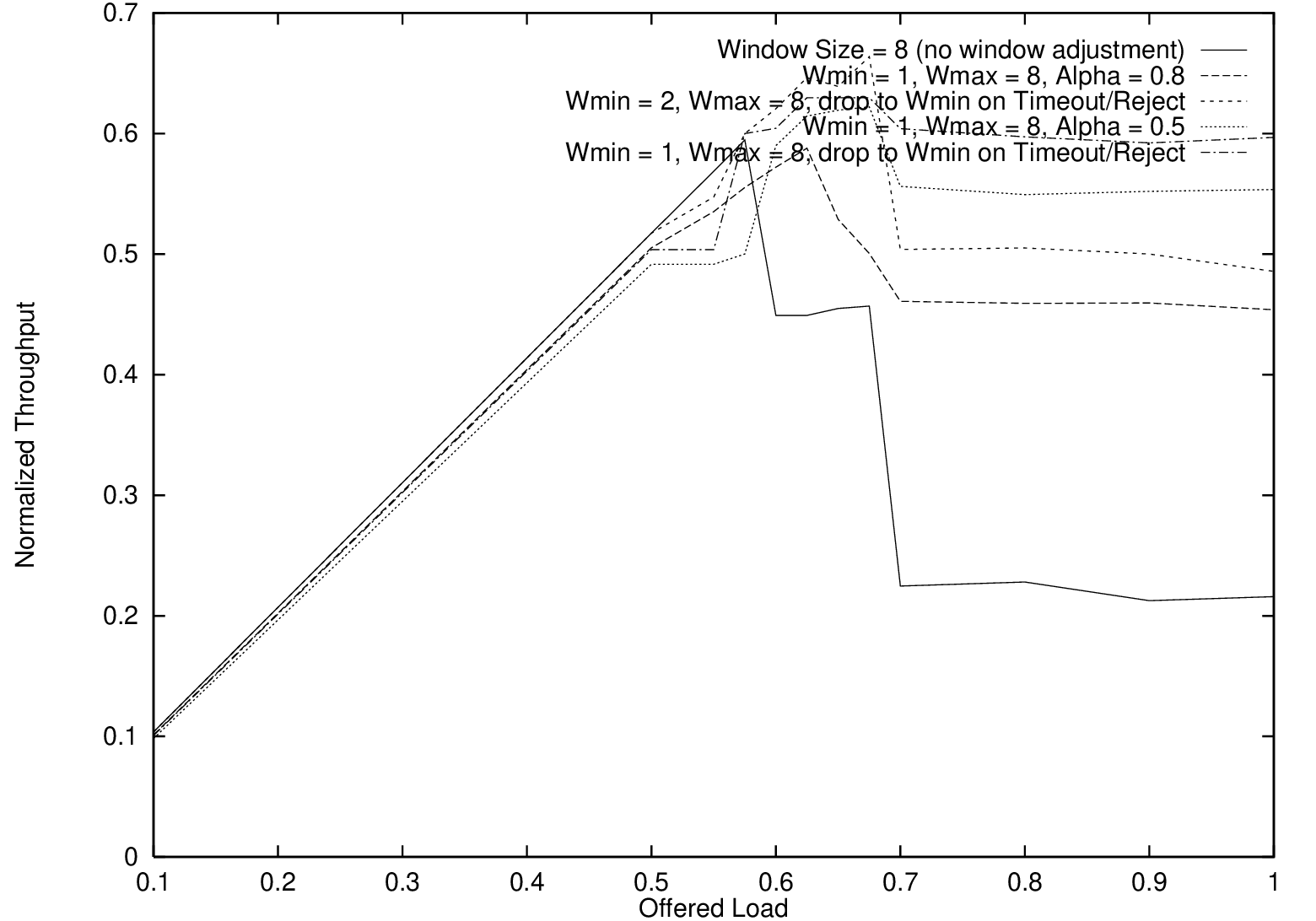,width=4.0in}}
        \caption{Variable Window Throughput Comparison.}
        \label{varwin}
\end{figure*}

\subsection{Flow Control}

Also in \cite{Nelson:1987:SCT}, it is suggested that flow control, shown
in Figure \ref{varwin}, in a
communications network exhibits not only brittle behavior, but
catastrophic behavior. The specific model of flow control considered in
\cite{Nelson:1987:SCT} is to divide available buffer space into classes
and allow packets which have passed through $i$ hops to occupy buffers
assigned to class $i$.
\section{Techniques for Handling Brittle Systems}

There are a variety of techniques for controlling and enhancing the 
ductility of a system. The first is to assign values to design
parameters which influence ductility in a static manner, that is
before the system becomes operational.
The next involves dynamically changing the ductility as the system
operates. This would be analogous to a material which could automatically 
trade-off hardness for ductility whenever necessary in order to maximize
its performance. The remaining techniques involve methods of attempting to 
avoid brittle fracture, by design or by rolling back from a fracture.

\subsection{Ductility Setting of System Sub-Components}

Now that ductility has been defined and the design parameters
controlling ductility identified, a natural question to ask is how should the
sub-component parameters be set. Within normal operation, the
performance requirements must be met, and in addition we would like the
system to be tough (robust) outside the normal operating range as well.
Is there a benefit to how ductility is distributed among subsystem
components? As an example, in network and transport level
data communications systems, if the system is going to fail, it is beneficial
for low level system components to fail early in the transmission
process rather than transporting a packet close to its destination and
finding that the entire packet/frame has to be retransmitted later.
Thus, it would be better to set $X_1$, in Figure \ref{bnet}, so that 
sub-component $S1$, which performs its processing early, has a lower 
ductility than components later in the process. 

\begin{figure*}[htbp]
        \centerline{\psfig{file=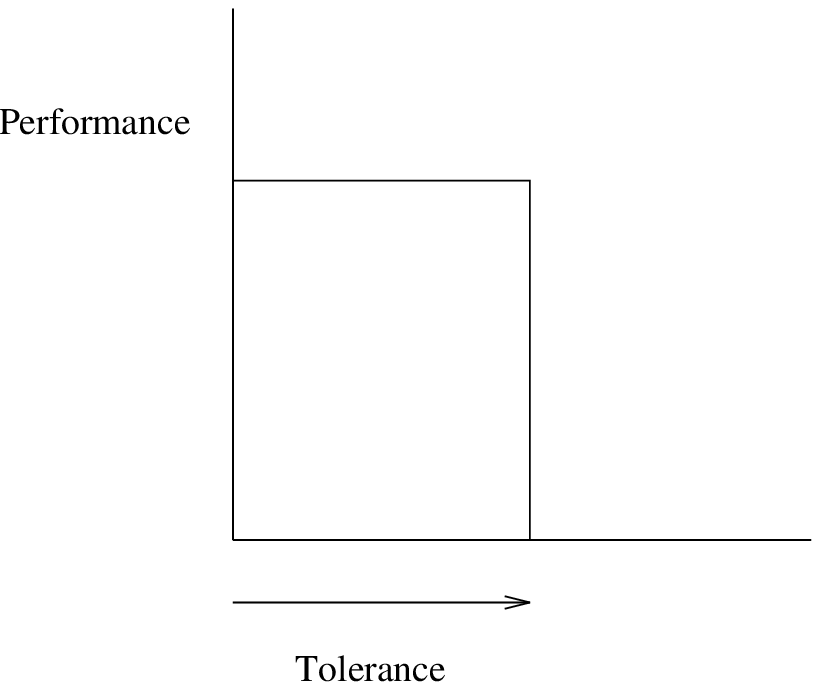,width=2.5in}}
        \caption{An On-Off Brittle Sub-Component.}
        \label{onoff}
\end{figure*}

A highly brittle component, as illustrated in Figure \ref{onoff}, would
appear to have the characteristics of an on-off constant bit rate
(on-off CBR) source. These types of sources have been used to model ATM
\cite{Prycker} traffic sources. Queue fill distribution 
has been analyzed in \cite{Anick1982} for on-off CBR models. These results
could be used in a buffer solution for such highly brittle components.

\subsection{Adaptation}

As mentioned previously, there are two forms of strain, reversible 
and permanent. Reversible
strain is called elastic strain and is characterized by Young's modulus:
the ratio of the stress over the strain. Permanent strain leaves the
shape of a material permanently changed and is known as plastic strain.
In brittle systems, an analog to plastic strain is adaptation.
Once we know the parameters which affect ductility, that is,
having determined $\psi$, the
values of the parameters can be changed dynamically as stress causes the
system to approach a brittle fracture.

\subsection{Rollback}

Another possibility is that once the system approaches a brittle 
fracture, the system has the capability to rollback to a safe state
and choose another gradient which attempts to remain
in a safe state of operation. Rollback techniques within a
communications network environment have been described in
\cite{BushThesis}.
%
 \acrodef{A}{Anti-Message}\index{Anti-Message}
 \acrodef{ABR}{Available Bit Rate}\index{Available Bit Rate}
 \acrodef{AHDLC}{Adaptive High Level Data Link Protocol}\index{Adaptive High Level Data Link Protocol}
 \acrodef{AMPS}{Advanced Mobile Phone System}\index{Advanced Mobile Phone System}
 \acrodef{ANO}{Attach New to Old}\index{Attach New to Old}
 \acrodef{ARP}{Address Resolution Protocol}\index{Address Resolution Protocol}
 \acrodef{ATMARP}{Asynchronous Transfer Mode Address Resolution Protocol}\index{Asynchronous Transfer Mode Address Resolution Protocol}
 \acrodef{ATM}{Asynchronous Transfer Mode}\index{Asynchronous Transfer Mode}
 \acrodef{AVNMP}{Active Virtual Network Management Prediction}
 \acrodef{BTS}{Base Transceiver Station}\index{Base Transceiver Station}
 \acrodef{C/E}{Condition Event Network}\index{Condition Event Network}
 \acrodef{CDMA}{Code Division Multiple Access}\index{Code Division Multiple Access}
 \acrodef{CDPD}{Cellular Digital Packet Data}\index{Cellular Digital Packet Data}
 \acrodef{CE}{Clustered Environment}\index{Clustered Environment}
 \acrodef{CMB}{Chandy-Misra-Bryant}\index{Chandy-Misra-Bryant}
 \acrodef{CMIP}{Common Management Information Protocol}\index{Common Management Information Protocol}
 \acrodef{CRC}{Cyclic Redundancy Checksum}\index{Cyclic Redundancy Checksum}
 \acrodef{CSS}{Cell Site Switch}\index{Cell Site Switch}
 \acrodef{CS}{Current State}\index{Current State}
 \acrodef{CTW}{Clustered Time Warp}\index{Clustered Time Warp}
 \acrodef{DES}{Data Encryption Standard}\index{Data Encryption Standard}
 \acrodef{DHCP}{Dynamic Host Configuration Protocol}\index{Dynamic Host Configuration Protocol}
 \acrodef{EN}{Edge Node}\index{Edge Node}
 \acrodef{ES}{Edge Switch}\index{Edge Switch}
 \acrodef{ESN}{Electronic Serial Number}\index{Electronic Serial Number}
 \acrodef{FH}{Fixed Host}\index{Fixed Host}
 \acrodef{FSM}{Finite State Machine}\index{Finite State Machine}
 \acrodef{GFC}{Generic Flow Control}\index{Generic Flow Control}
 \acrodef{GPS}{Global Positioning System}\index{Global Positioning System}
 \acrodef{GSM}{Global System for Mobile Communication}\index{Global System for Mobile Communication}
 \acrodef{GSV}{Global Synchronic Distance}\index{Global Synchronic Distance}
 \acrodef{GVT}{Global Virtual Time}\index{Global Virtual Time}
 \acrodef{HDLC}{High Level Data Link Protocol}\index{High Level Data Link Protocol}
 \acrodef{HLR}{Home Location Register}\index{Home Location Register}
 \acrodef{HO}{Handoff}\index{Handoff}
 \acrodef{IETF}{Internet Engineering Task Force}\index{Internet Engineering Task Force}
 \acrodef{IPC}{Inter-Processor Communication}\index{Inter-Processor Communication}
 \acrodef{LIS}{Logical IP Subnet}\index{Logical IP Subnet}
 \acrodef{LN}{Logical Node}\index{Logical Node}
 \acrodef{LP}{Logical Process}\index{Logical Process}
 \acrodef{LVT}{Local Virtual Time}\index{Local Virtual Time}
 \acrodef{MAC}{Media Access Control}\index{Media Access Control}
 \acrodef{MH}{Mobile Host}\index{Mobile Host}
 \acrodef{MIB}{Management Information Base}\index{Management Information Base}
 \acrodef{MIN}{Mobile Identification Number}\index{Mobile Identification Number}
 \acrodef{MSC}{Mobile Switching Center}\index{Mobile Switching Center}
 \acrodef{MSR}{Mobile Support Router}\index{Mobile Support Router}
 \acrodef{MTW}{Moving Time Windows}\index{Moving Time Windows}
 \acrodef{MT}{Mobile Terminal}\index{Mobile Terminal}
 \acrodef{NCP}{Network Control Protocol}\index{Network Control Protocol}
 \acrodef{NFT}{No False Time-stamps}\index{No False Time-stamps}
 \acrodef{NIPAT}{Network Insecurity Path Assessment Tool}
 \acrodef{NHRP}{Next Hop Resolution Protocol}\index{Next Hop Resolution Protocol}
 \acrodef{NPSI}{Near Perfect State Information}\index{Near Perfect State Information}
 \acrodef{NTP}{Network Time Protocol}\index{Network Time Protocol}
 \acrodef{PA}{Perturbation Analysis}\index{Perturbation Analysis}
 \acrodef{PBS}{Portable Base Station}\index{Portable Base Station}
 \acrodef{PCN}{Personal Communications Network}\index{Personal Communications Network}
 \acrodef{PDES}{Parallel Discrete Event Simulation}\index{Parallel Discrete Event Simulation}
 \acrodef{PDU}{Protocol Data Unit}\index{Protocol Data Unit}
 \acrodef{PGL}{Peer Group Leader}\index{Peer Group Leader}
 \acrodef{PG}{Peer Group}\index{Peer Group}
 \acrodef{P/T}{Place Transition Net}\index{Place Transition Net}
 \acrodef{PIPS}{Partially Implemented Performance Specification}\index{Partially Implemented Performance Specification}
 \acrodef{PNNI}{Private Network-Network Interface}\index{Private Network-Network Interface}
 \acrodef{PP}{Physical Process}\index{Physical Process}
 \acrodef{Q.2931}{Q.2931}\index{Q.2931}
 \acrodef{QR}{Receive Queue}\index{Receive Queue}
 \acrodef{QS}{Send Queue}\index{Send Queue}
 \acrodef{QoS}{Quality of Service}\index{Quality of Service}
 \acrodef{RDRN}{Rapidly Deployable Radio Network}\index{Rapidly Deployable Radio Network}
 \acrodef{RN}{Remote Node}\index{Remote Node}
 \acrodef{RT}{Real Time}\index{Real Time}
 \acrodef{SID}{Station Identification}\index{Station Identification}
 \acrodef{SLP}{Single Processor Logical Process}\index{Single Processor Logical Process}
 \acrodef{SLW}{Sliding Lookahead Window}\index{Sliding Lookahead Window}
 \acrodef{SNMP}{Simple Network Management Protocol}\index{Simple Network Management Protocol}
 \acrodef{SQ}{State Queue}\index{State Queue}
 \acrodef{SS7}{Signaling System 7}\index{Signaling System 7}
 \acrodef{TDMA}{Time Division Multiple Access}\index{Time Division Multiple Access}
 \acrodef{TDN}{Temporary Directory Number}\index{Temporary Directory Number}
 \acrodef{TNC}{Terminal Node Controller}\index{Terminal Node Controller}
 \acrodef{TOE}{Time of Expiry}\index{Time of Expiry}
 \acrodef{TR}{Receive Time}\index{Receive Time}
 \acrodef{TS}{Send Time}\index{Send Time}
 \acrodef{VC}{Virtual Circuit}\index{Virtual Circuit}
 \acrodef{VCI}{Virtual Circuit Identifier}\index{Virtual Circuit Identifier}
 \acrodef{VLR}{Visiting Location Register}\index{Visiting Location Register}
 \acrodef{VNC}{Virtual Network Configuration}\index{Virtual Network Configuration}
 \acrodef{VP}{Virtual Path}\index{Virtual Path}
 \acrodef{VPI}{Virtual Path Identifier}\index{Virtual Path Identifier}
 \acrodef{VTRP}{Virtual Trees Routing Protocol}\index{Virtual Trees Routing Protocol}
%
%
\medskip
\bibliographystyle{IEEE} 
\bibliography{/home/bushsf/ref/mob,/home/bushsf/ref/twe,/home/bushsf/ref/psim,/home/bushsf/ref/atmmob,/home/bushsf/ref/standards,/home/bushsf/ref/vci,/home/bushsf/ref/handoff,/home/bushsf/ref/theorem_proving,/home/bushsf/ref/topo,/home/bushsf/ref/signaling,/home/bushsf/ref/an,/home/bushsf/ref/iw,/home/bushsf/ref/brittle}
\begin{biography}{Stephen F. Bush} 
Stephen F. Bush is a Computer Scientist at General Electric Research
and Development (GE CR \& D) in Niskayuna, NY. Steve is currently the
Principal Investigator for the DARPA funded Active Networks Project at GE. 
Before joining GE CR \& D, Stephen was a researcher at the Information and
Telecommunications Technologies Center (ITTC) at the University of Kansas
where he contributed to the DARPA Rapidly Deployable Radio Networks Project.
He received his B.S. in Electrical and Computer Engineering from
Carnegie Mellon University and M.S. in Computer Science from Cleveland
State University. He has worked many years for industry in the areas of
computer integrated manufacturing and factory automation and control.
Steve received the award of Achievement for Professional Initiative and 
Performance for his work as Technical Project Leader at GE Information 
Systems in the areas of network management and control while pursuing
his Ph.D. at Case Western Reserve University. Steve completed his Ph.D.
research at the University of Kansas where he received a Strobel Scholarship
Award. He can be reached at bushsf@crd.ge.com and 
http://www.crd.ge.com/people/bush.
\end{biography}

\end{document}